\documentclass{ws-procs9x6}

\begin{document}

\title{QCD AT COMPLEX COUPLING, LARGE ORDER IN PERTURBATION THEORY AND THE GLUON CONDENSATE}

\author{Y. MEURICE}

\address{Department of Physics and Astronomy, University of Iowa,\\
Iowa City, IA 52242, USA\\
$^*$E-mail: yannick-meurice@uiowa.edu\\}

\begin{abstract}
We discuss the relationship between the large order behavior of the perturbative  series for the average plaquette in pure gauge theory and singularities in the complex coupling plane. We discuss simple extrapolations of the large order for this series. 
We point out that when these extrapolated series are subtracted from the Monte Carlo 
data, one obtains (naive) estimates of the gluon condensate that are significantly 
larger than values commonly used in the continuum for phenomelogical purpose.
We present numerical results concerning the zeros of the partition function in the complex 
coupling plane (Fisher's zeros). We report recent attempts to solve this problem using 
the density of states. We show that weak and strong coupling expansions for the 
density of states agree surprisingly well for values of the action relevant for the 
crossover regime.

\end{abstract}


\bodymatter

\section{Motivations}
Perturbation theory has played a major role in the establishment of the standard model 
of electro-weak and strong interactions. However, it is well known \cite{dyson52}
that perturbative series have a zero radius of convergence and that 
at some critical order, 
adding more terms does not improve the accuracy of the result. 

There exists a connection between large field configurations and the large order 
of perturbative series \cite{convpert} that can be illustrated with the simple integral
\begin{equation}
\int_{-\infty}^{+\infty}d\phi e^{-\frac{1}{2}\phi^2-\lambda \phi^4}\neq \sum_0^{\infty}
\frac{(-\lambda)^l}{l!} \int_{-\infty}^{+\infty}d\phi e^{-\frac{1}{2}\phi^2}\phi^{4l}
\end{equation}
The truncation of the exponential at order $l$ is justified if the argument is much smaller than $l$. However, the peak of the integrand is located at $\phi=\sqrt{4l}$. For this value of $\phi$, the argument of the exponential is $\lambda 16 l^2$, which for $l$ large enough will be larger than $l$. In other words, 
the peak of the integrand of the r.h.s. moves too fast when the order increases.
On the other hand, if we introduce a field cutoff, the peak moves outside of the integration range and 
\begin{equation}
\int_{-\phi_{max}}^{+\phi_{max}}d\phi e^{-\frac{1}{2}\phi^2-\lambda \phi^4}= \sum_0^{\infty}
\frac{(-\lambda)^l}{l!} \int _{-\phi_{max}}^{+\phi_{max}}d\phi e^{-\frac{1}{2}\phi^2}\phi^{4l}
\end{equation}
This example suggests that one should use perturbation theory to treat small quantum 
fluctuations and semi-classical methods to treat the large field configurations. 

For QCD, an important challenge is to describe the large distance behavior of the 
theory in terms of  degrees of freedom which are weakly coupled at short distance. 
This question can be addressed in the framework of the lattice formulation. 
We consider the simplest case of Wilson's action which is  the sum over the plaquettes in 
the fundamental $SU(N)$ representation:
\begin{equation} 
S=\sum_{p}(1-(1/N)Re Tr(U_p))\  .
\end{equation}
With the usual notation 
$\beta=2N/g^2$, the partition function reads
\begin{equation}
Z=\prod_{l}\int dU_l {\rm e}^{-\beta S} \  .
\end{equation}
For $N=3$, this theory has no phase transition when one goes from small coupling 
(large $\beta$) to large coupling (small $\beta$).  Recently, convincing argument have 
been given \cite{tomboulis07,tomboulis07b} 
in favor of the smoothness of the RG flows between the two corresponding fixed points. 
Consequently, there does not seem to exist any fundamental obstruction 
to match the two regimes.  One general question that we would like to address is if it is possible to construct a modified weak coupling expansion that could bridge the gap 
to the strong coupling regime. 

Lattice gauge theories with a {\it compact} group have a build-in large field cutoff: the group elements associated with the links are integrated with 
$dU_l$, the compact Haar measure. For $SU(2)$ and $SU(3)$, 
the action density has an upper bound which is saturated when the group element is a 
nontrivial element of the center. It is remarkable that this formulation has 
a UV regularization {\it and} a large field regularization that preserves gauge invariance.
Does the presence of large field cutoff means that perturbative series are convergent?
Not necessarilly, because in constructing perturbative series, one decompactifies \cite{heller84} the gauge field integration which at low order amounts to neglect exponentially small tails of integration, but modifies the large order drastically.  

In continuous field theories, 
it is expected \cite{bender69,Parisi77,brezin77,leguillou90} that the large order of perturbative series can be calculated 
from classical solutions at small negative $\lambda$ for scalar theories or small negative $g^2$ for gauge theories. For lattice gauge theory with compact groups, 
the theory is mathematically well defined at negative $g^2$ ,i. e., negative $\beta$, 
but  some Wilson loops oscillate when their size change and the average plaquette has a discontinuity when $g^2$ changes sign\cite{gluodyn04}.
For small negative $g^2$,  i. e., large negative $\beta$, the behavior 
of expectation values is dominated by the large field configurations. This 
statement can be made more precise by introducing the spectral decomposition
\begin{equation}
Z=\int_0^{S_{max}}dS\ n(S){\rm e}^{-\beta S}\  ,
\end{equation}
with $n(S)$ the so-called density of states. It is clear that for large negative $\beta$, 
what matters is the behavior of $n(S)$ near $S_{max}$. 

We can make the discussion more concrete by considering the case of the single plaquette 
model \cite{plaquette} with $SU(2)$ gauge group. In this case, $S_{max}=2$ and we have 
 \begin{equation}
 n_{1 pl.}(S)=\frac{2}{\pi}\sqrt{S(2-S)}\  .
 \end{equation} 
The weak coupling expansion of $Z$ is determined by the behavior of near $S=2$.
When we expand 
$\sqrt{2-S}$ about $S=0$, the large order of this series is determined by the cut at $S=2$. After integration (from 0 to $+\infty$) over $S$, the series with a finite radius 
of convergence becomes asymptotic (inverse Borel transform). Alternatively, we can 
maintain the finite range of integration and construct a converging weak coupling 
expansion but with coupling dependent coefficients as done by Hadamard a century 
ago in his study of Bessel functions. It would be interesting to know if the features of the one plaquette model persist in the infinite volume limit of lattice gauge theory.

In these proceedings, we discuss the large order behavior of the weak coupling expansion of the  plaquette (Sec. 2) and 
the possibility of defining its non-perturbative part (the gluon condensate? Sec. 3) .
More details can be found in Ref. \cite{npp}. 
In Sec. 4, we present numerical results concerning the zeros of the partition function in the complex 
coupling plane  \cite{quasig,lat07} (Fisher's zeros). This problem could be solved using 
the density of states. Preliminary numerical results concerning the density of states 
are provided in Sec. 5 where we also show that weak and strong coupling expansions for the 
density of states agree surprisingly well for values of the actions relevant for the 
crossover regime. After the conference, we wrote a more detailed preprint 
\cite{Denbleyker:2008ss} concerning this question.
\section{Perturbative series in lattice gauge theory}
In this section, we denote $\mathcal{N}_p\equiv\ L^D D(D-1)/2\ $  the 
number of plaquettes and the 
average plaquette: 
\begin{equation}
P(\beta)=(1/\mathcal{N}_p)\left\langle \sum_p
(1-(1/N)Re Tr(U_p))\right\rangle \  .
\end{equation}
The weak coupling expansion of this quantity
\begin{equation}
P(\beta)\simeq \sum_{m=1}b_m \beta^{-m} +\dots.
\end{equation}
has been calculated up to order ten \cite{direnzo2000} and 16 \cite{rakow05}.
Series analysis  
\cite{rakow2002,third}
suggest a singularity $P\propto (1/5.74-1/\beta)^{1.08}$ and consequently a finite 
radius of convergence. This is not expected  
since the plaquette 
changes discontinuously \cite{gluodyn04} at $\beta \rightarrow \pm \infty$. 
This behavior is also not seen in the 2d derivative of $P$ and would require massless glueballs. 

A simple alternative is that the critical 
point in the fundamental-adjoint plane \cite{bhanot81} has mean field exponents and in particular $\alpha=0$. 
On the $\beta_{adj.}=0$ line, we 
assume an approximate logarithmic behavior (mean field)
\begin{equation}
-\partial P/\partial \beta \propto {\rm ln}((1/\beta_m-1/\beta)^2+\Gamma^2)\  .
\end{equation}
$\Gamma$ cannot be too small (absence of singularities) or too large (this 
would create visible oscllations in the perturbative coefficients). 
From these constraints, we get the approximate bounds \cite{npp}
 $0.001<\Gamma<0.01$.

Integrating, we get the approximate form 
\begin{equation}
\sum _{k=0} b_k\beta^{-k}\simeq C({\rm Li}_2 (\beta^{-1}/(\beta_m^{-1}+i\Gamma))+{\rm h.c}\  ,
\end{equation}
with 
\begin{equation}
{\rm Li}_2(x)=\sum _{k=0}x^k/k^2 \  .
\end{equation}
We fixed $\Gamma=0.003$ and obtained $C=0.0654$ and $\beta_m$=5.787 using 
of $b_9$ and $b_{10}$. The low order coefficients depend very little on $\Gamma$ (when $\Gamma<0.01$), larger series are needed to get a better estimate of 
$\Gamma$ for $SU(3)$. It interesting to notice that 
we get very good predictions of the values of $b_8, b_7, \dots $!
\begin{table}
\tbl{Predicted values of $b_m$ with the dilog model.}
{\begin{tabular}{cccc}\toprule
{\rm order} & {\rm predicted} & {\rm numerical}  \cite{direnzo2000} &{\rm rel. error}\\\colrule
 1 & \text{0.7567} & 2 &\text{-0.62}\\
 2 & \text{1.094} &
   \text{1.2208}&\text{-0.10} \\
 3 & \text{2.811} &
   \text{2.961} &\text{-0.05}\\
 4 & \text{9.138} &
   \text{9.417}&\text{-0.03}\\
 5 & \text{33.79} &
   \text{34.39}&\text{-0.017} \\
 6 & \text{135.5} &
   \text{136.8}&\text{ -0.009}\\
 7 & \text{575.1} &
   \text{577.4}&\text{-0.004} \\
 8 & \text{2541} & 2545& \text{-0.0016}\\
 9 & exact & 11590 \\
 10 & exact& 54160\\\botrule
\end{tabular}}
\end{table}
We believe that these regularities should have a Feynman diagram interpretation.

Another possible model \cite{mueller93, direnzo95} is based on IR renormalons 
\begin{equation}
\sum _{k=0} b_k\bar{\beta}^{-k} \simeq K \int_{t_1}^{t_2}dt {\rm e}^{-\bar{\beta}t}\ (1-t\ 33/16\pi^2)^{-1-204/121}
\end{equation}
\begin{equation}
\bar{\beta}=\beta(1+d_1/\beta+\dots)
\label{eq:shift}
\end{equation}
$t_1=0$ corresponds to the UV cutoff while 
$t_2=16\pi^2/33$ corresponds to the Landau pole. For $t_2 = \infty$ we get  a 
perturbative series with a factorial growth, in contrast with the previous model 
which had a power growth. Unfortunately a clear distinction between the two types 
of large order behavior requires numerical calculations at order larger than 20. 

\section{The  Gluon Condensate}
Using the two large order extrapolations described in the previous section, 
one can see good evidence \cite{npp} for 
\begin{equation}
P(\beta)-P_{pert.}(\beta) \simeq B (a/r_0)^4\  ,
\end{equation}
with $a(\beta)$ defined with the so-called force  scale, \cite{guagnelli98,necco01}  and $P_{pert}$ appropriately truncated for the second model.
$B$ is sensitive to resummation. $B\simeq 0.7$ with the bare series \cite{npp}
and 0.4 with the tadpole improved series \cite{rakow05}. 

It is tempting but potentially dangerous to try to relate $B$ to the numerical value of the 
gluon condensate\cite{Shifman:1978bx}. If we identify \cite{digiacomo81} for $N=3$
\begin{equation}
P(\beta)-P_{pert.}(\beta) \simeq a^4\frac{\pi^2}{36}<\frac{\alpha}{\pi}GG> ,
\end{equation}
we obtain for $r_0=0.5\ fm$ and $B=0.4$ that $<\frac{\alpha}{\pi}GG> \simeq 0.035 \ GeV^4$ which 
is about 3 times the original estimate\cite{Shifman:1978bx}.  It is not clear to us that 
there is a precise correspondence between the continuum and the lattice definitions. 
Also, different values for the continuum value have been proposed. 
A negative value \cite{Davier:2008sk}, subsequently criticized \cite{Maltman:2008nf} can even be found. For these reasons, some authors \cite{Davies:2008sw} prefer to use 
a value 0 with error bars when estimating $\alpha_s$ at some large scale. 
If we use the correspondence between the lattice and the continuum discussed above at face value with 
$B=0.4-0.7$, these error bars should be multiplied by a factor 3 to 6. 

\section{ Zeros of the Partition Function}
\begin{figure}
\begin{center}
\includegraphics[width=3.in,angle=270]{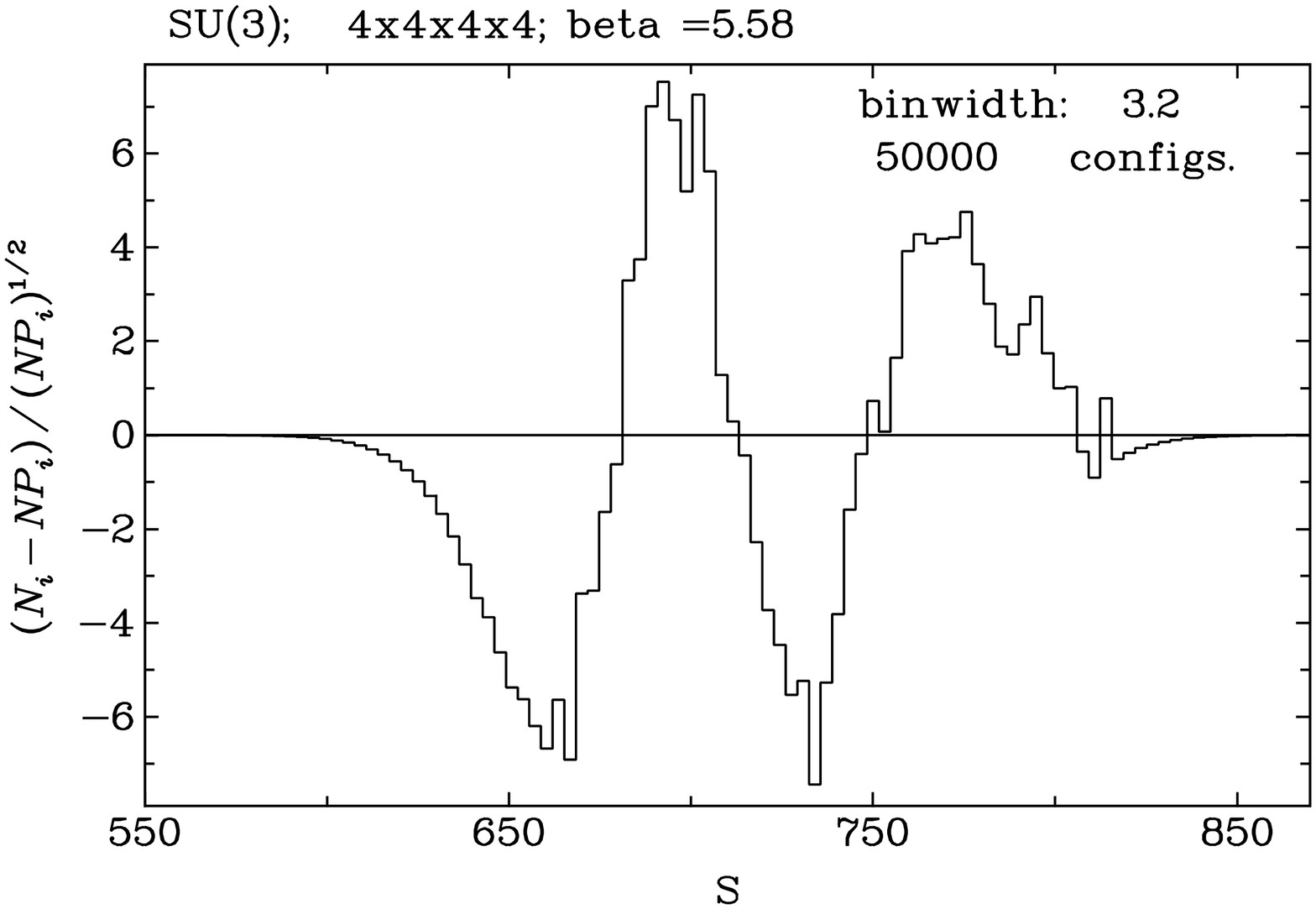}
\includegraphics[width=3.in,angle=270]{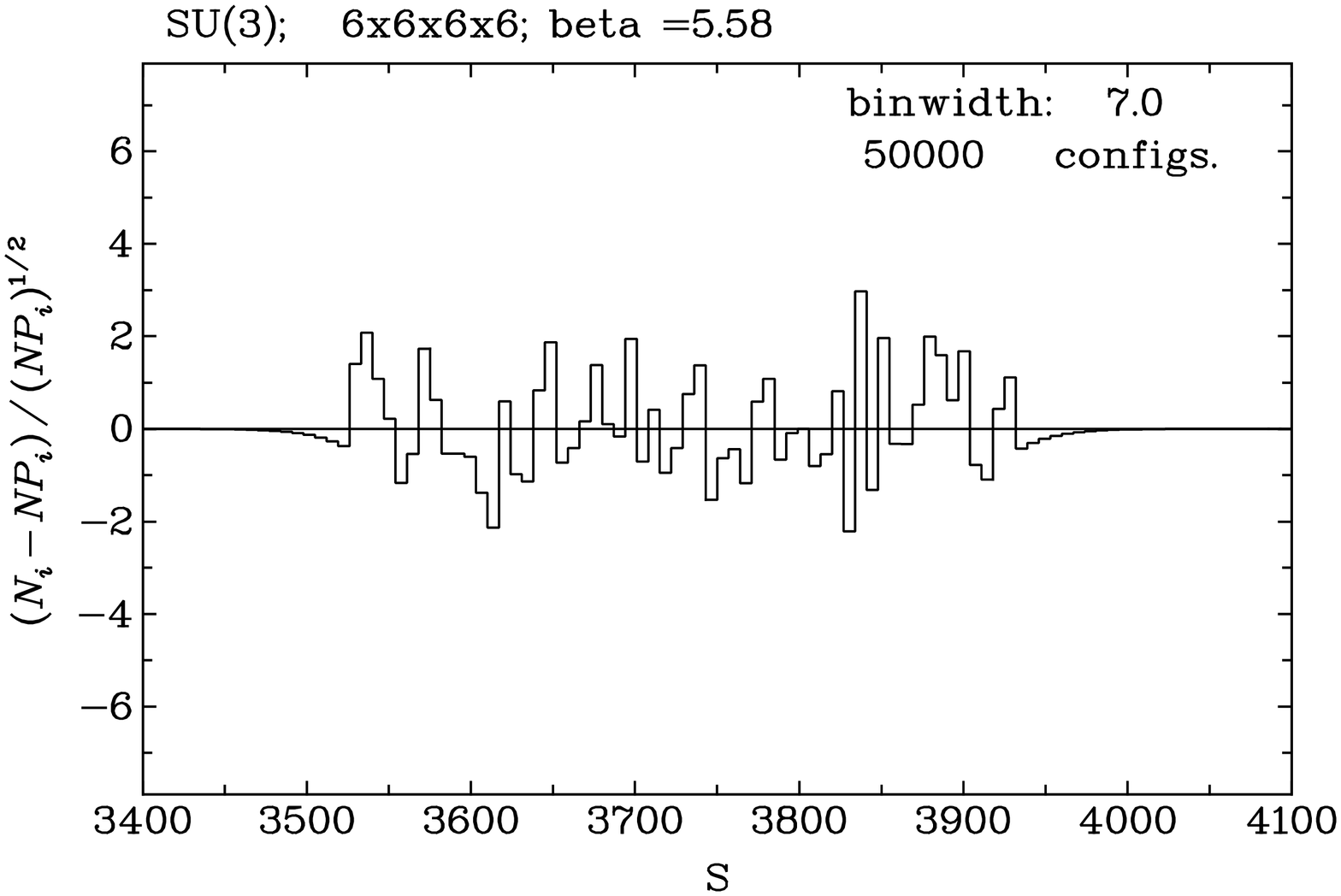}
\caption{The relative difference with a Gaussian distribution for 50,000 values of $S$ in an histogram with 100 bins 
for a $SU(3)$ pure gauge theory on $4^4$ and $6^4$ lattices at $\beta =5.58$ .  }
\end{center}
\label{fig:resid}
\end{figure}

\begin{figure}
\begin{center}
\includegraphics[width=3.in,angle=270]{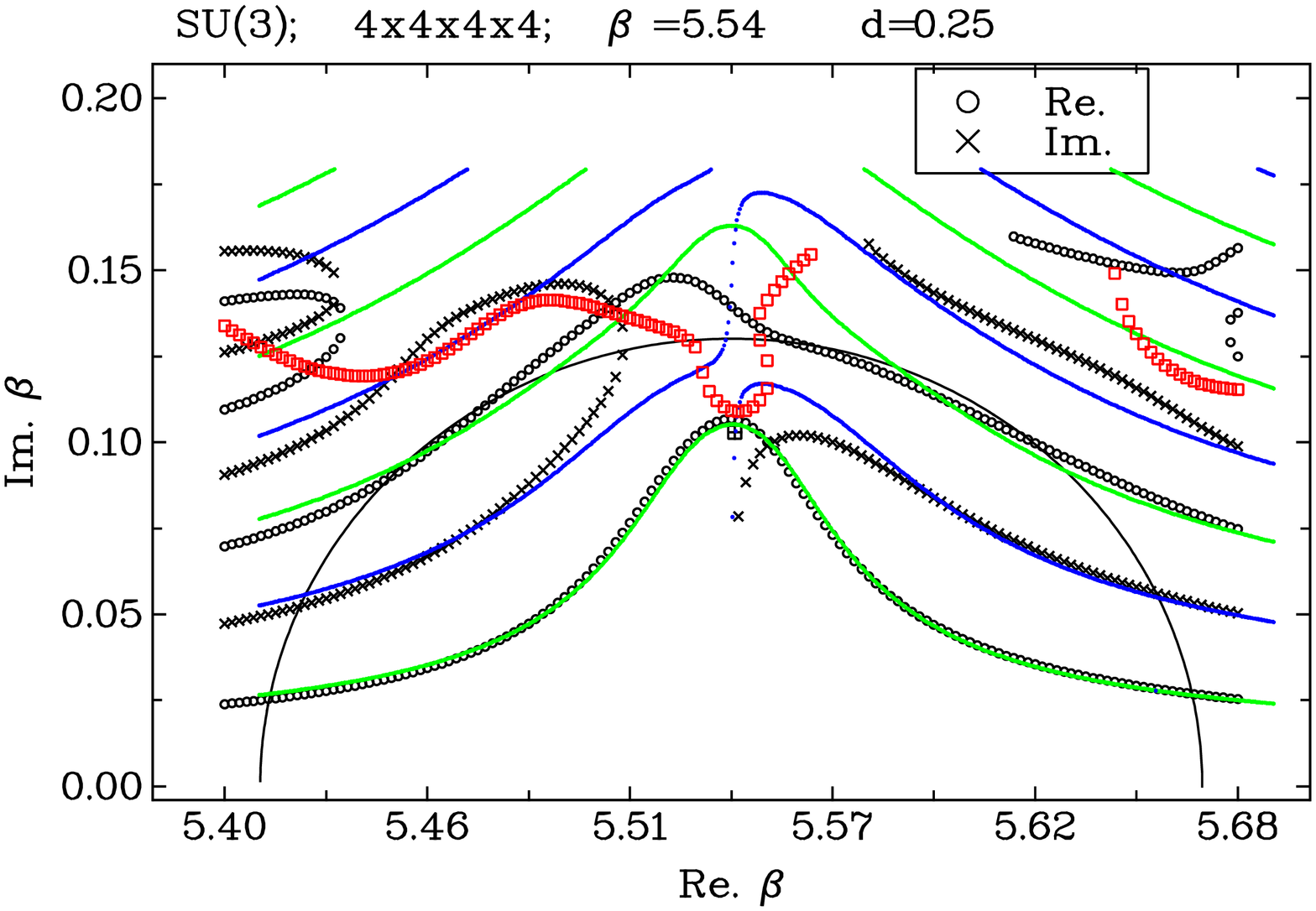}
\includegraphics[width=3.in,angle=270]{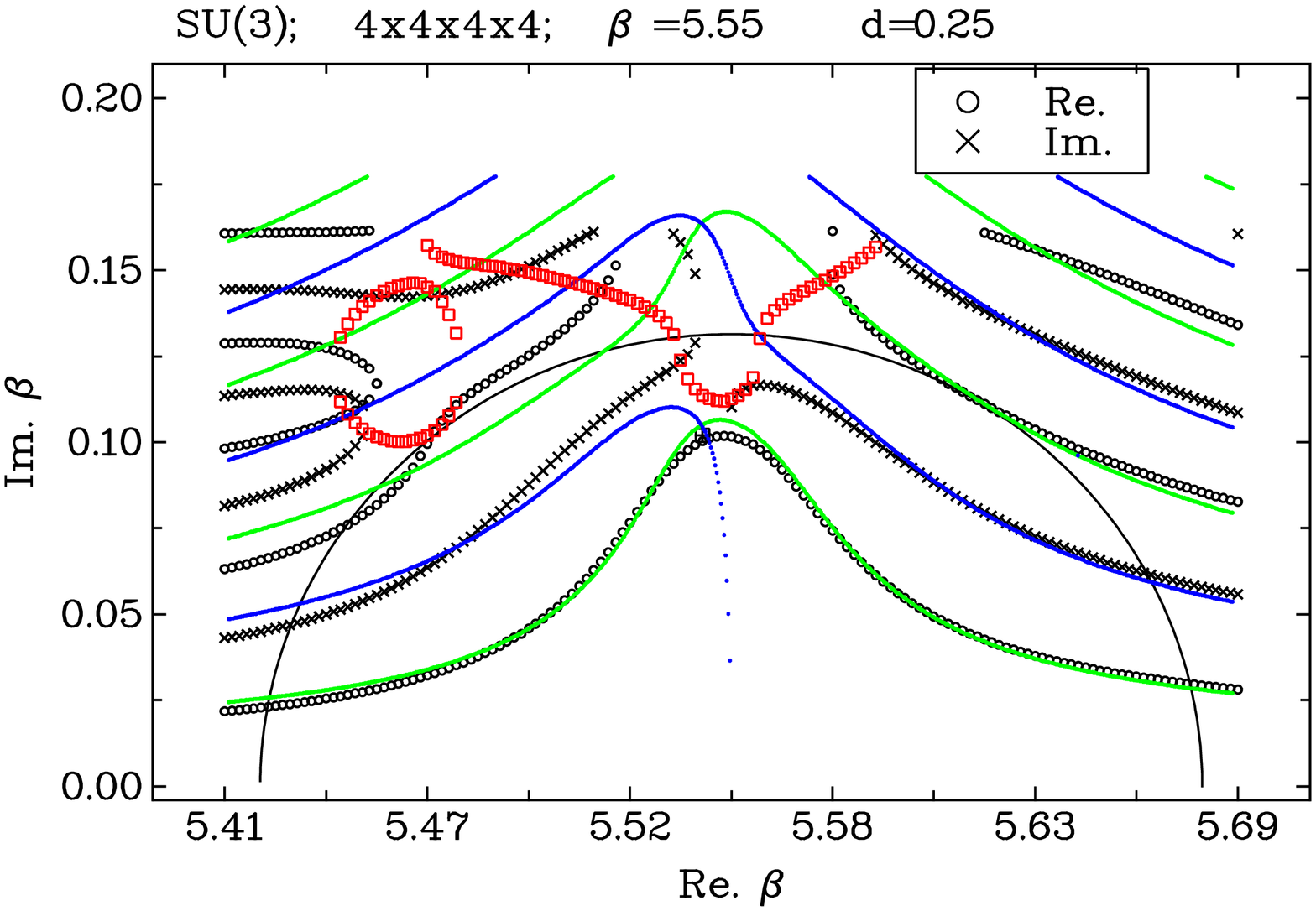}
\caption{{\label{fig:su3z}} Zeros of the real (circles) and imaginary (crosses) 
for $SU(3)$ on a $4^4$ lattice using reweighting of MC data at $\beta_0$ = 5.54 and 5.55. The small dots are the 
values for the real (green) and imaginary (blue) parts 
obtained from the 4 parameter model. 
The crossed box at ((5.541,0.103) above and (5.542,0.102) below) have been obtained with a perturbative method. 
Red boxes: boundary of the MC confidence region. The solid line is the circle of confidence of the Gaussian approximation. The locations of the complex zeros are consistent.}
\end{center}
\end{figure}

The existence of complex singularities near the real axis can be tested by studying the 
complex zeros of the partition function \cite{fisher65}.
The basic technique is the reweighting \cite{falcioni82,alves90b} 
of action distributions at given $\beta_0$
\begin{equation}
Z(\beta_0+\Delta \beta)=Z(\beta_0)<\exp (-\Delta \beta S)>_{\beta_0}\ .
	\label{eq:pf}
\end{equation}
As mentioned in the introduction, $Z(\beta)$ is the Laplace transform of the  density of states $n(S)$. 
For $SU(2)$ with even numbers of sites in each direction \cite{gluodyn04}
$Z(-\beta)={\rm e}^{2\beta\mathcal{N}_p}Z(\beta)$ and $n(S)=n(2\mathcal{N}_p-S)$. In the crossover, we have attempted local parametrizations \cite{quasig,lat07}:
\begin{equation}
n(S)\propto {\rm e}^{-(a_1S+a_2S^2+a_3S^4+a_4S^4)}
\end{equation}
For $SU(3)$ on a $4^4$ lattice, stronger deviations from the Gaussian behavior are observed than for $SU(2)$. 
This is illustrated in Fig. \ref{fig:resid} that has a larger scale than its $SU(2)$ counterpart. 
The histogram were made with 50,000 configurations prepared for a study of the third and fourth 
moments \cite{third}. 
As the volume increases, these features tend to disappear in the statistical noise. 
These local model give results that can be compared to MC reweighting in the 
region where the errors on the zero level curves for the real and imaginary parts 
are not too large \cite{quasig}. An important consistency test is to show that approximately the same zeros are obtained for different $\beta_0$. This is 
illustrated in Fig. \ref{fig:su3z}. We plan to pursue this study using the more 
global information provided by the density of states.
\vfill
\eject
\section{ Approximate form of the density of state $n(S)$ for $SU(2)$}

We  assume the following form:
\begin{equation}
n(S)={\rm e}^{\mathcal{N}_p f(S/\mathcal{N}_p)}\ .
\end{equation}
In the infinite volume limit, 
$f(x)$ becomes volume independent and can be interpreted as a (color) entropy density.
In the same limit, we have the 
saddle point equation
\begin{equation}
f'(x)=\beta \ .
\end{equation}
This is the analog of the familiar thermodynamical relation  $dS/dE=1/T$.
Knowing $f(x)$ amounts to solve the theory (in a thermodynamical sense).
For $SU(2)$, $f(x)=f(2-x)$ (symmetric about 1) and we don't need to calculate 
$f$ for $x>1$. 
We have constructed weak coupling and strong coupling expansions for $SU(2)$ and 
compared with numerical data on a $4^4$ lattice. 

For the weak coupling, we use the large beta expansion near $x=0$, 
$x=<S/\mathcal{N}_p>= \frac{3}{4} \frac{1}{\beta} + 0.156 \frac{1}{\beta^2} + \dots $.  
We assume $f(x)\simeq A \ ln(x) + B +Cx+\dots$, plug the expansion in the saddle point equation and solve for 
$A$ and other unknowns. We obtained 
$f(x)=\frac{3}{4} ln(x) +  0.208 x + 0.0804 x^2 +\dots $.
For the strong coupling,  
we use the low beta expansion \cite{balian74}
to solve near $x=1$ for $SU(2)$, and obtained 
$x-1=<S/\mathcal{N}_p-1>=-\frac{\beta}{4}+\frac{\beta^3}{96}
-\frac{7\beta^5}{1536}  +\frac{31
   \beta^7}{23040}+\dots + \frac{1826017873 \beta^{15}}{68491306598400}$. 
With periodic boundary conditions, the coefficients have no volume 
dependence for graphs with trivial topology (volume dependence should appear at order $\beta^{2L}$).
Solving the saddle point equation for an expansion about 1, we get  
$f(x)=-2(x-1)^2-\frac{2}{3}(x-1)^4+\dots + \frac{163150033 (x-1)^{16}}{255150}+\dots$.

\begin{figure}
\vskip-30pt
\centerline{\psfig{figure=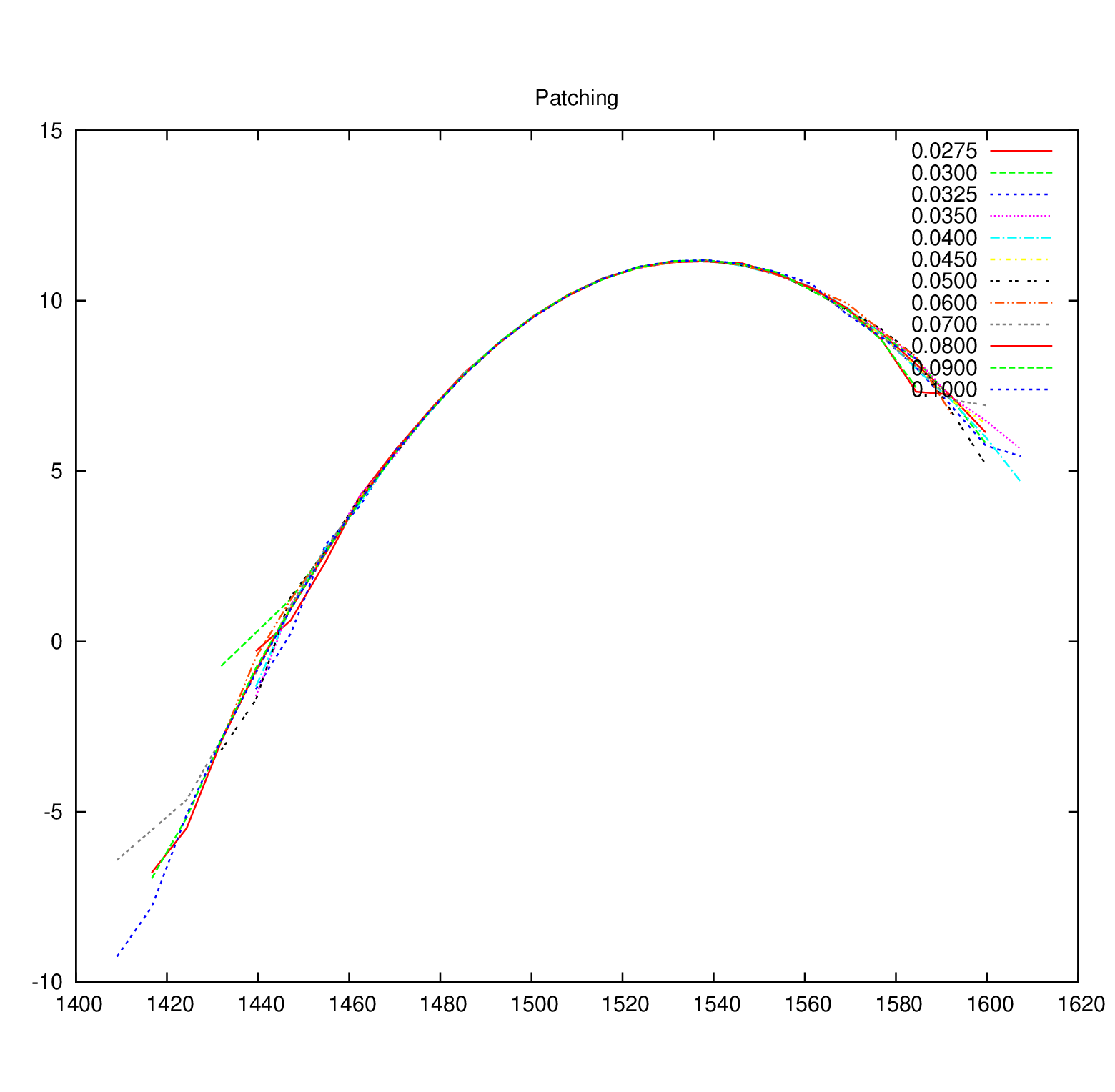,width=3.in,angle=270}}
\centerline{\psfig{figure=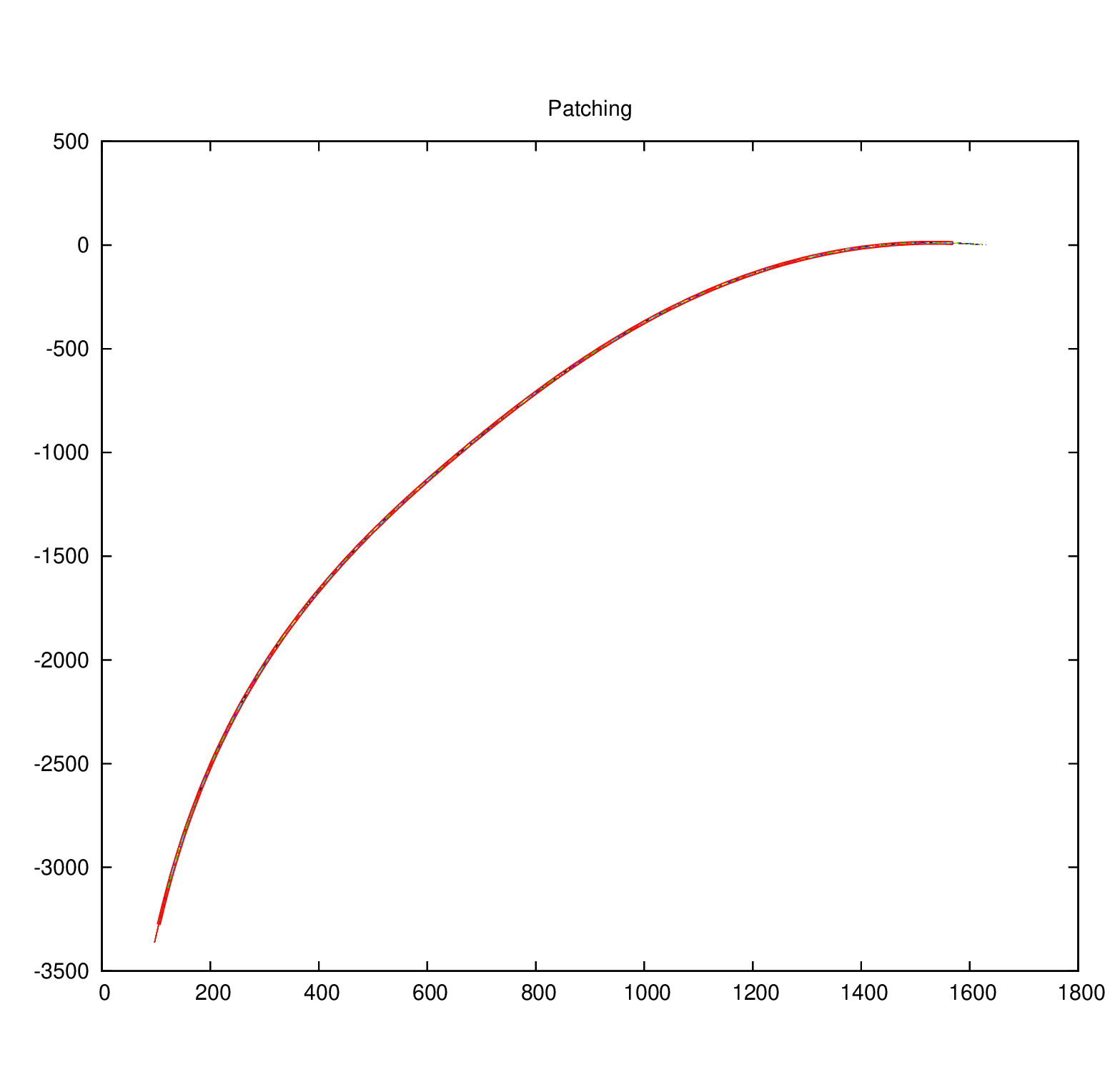,width=3.in,angle=270}}
\caption{\label{fig:extraps} $n(S) \propto  P_{\beta_0}(S){\rm e}^{\beta_0 S}$ for different $\beta_0$ (patching).  Collection of overlapping data (A. Denbleyker).
}
\end{figure}

\begin{figure}
\vskip-30pt
\centerline{\psfig{figure=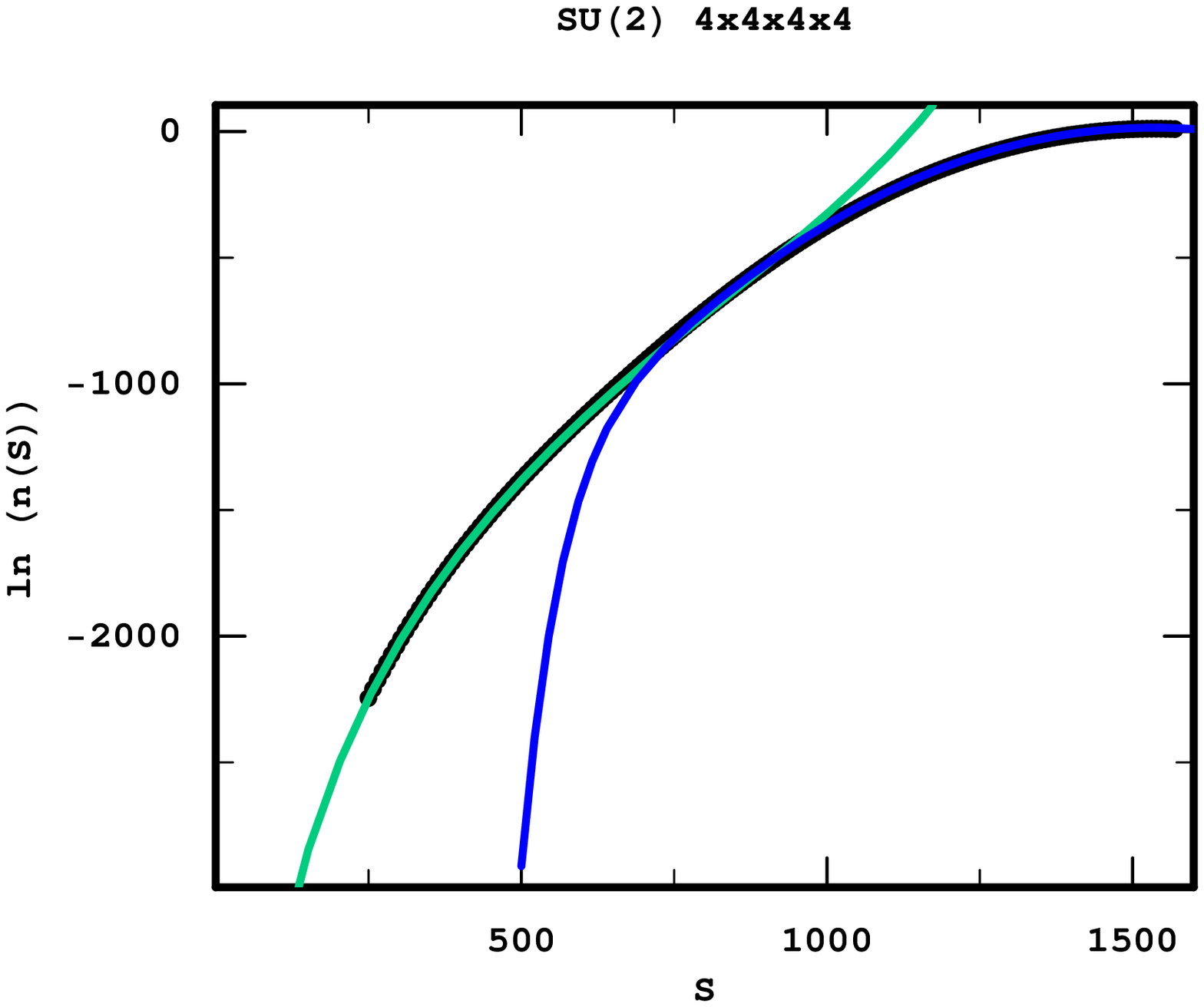,width=3.in,angle=0}}
\caption{\label{fig:compa}$ln(n(S))$ numerical (black), strong coupling at order 16 (blue) and weak coupling with dilog model at order 10 (green) for $SU(2)$ on a $4^4$ lattice.}
\end{figure}

The numerical construction of $n(S)$ by patching was done by A. Denbleyker and is illustrated in Fig. \ref{fig:extraps}. Series expansions of $f(x)$ are compared with the 
numerical data in Fig. \ref{fig:compa}. Note the good agreement in the intermediate region. 
After the conference, we wrote a more detailed preprint 
\cite{Denbleyker:2008ss} where details can be found.
\vfill
\eject
\section{Conclusions}

The density of state show a nice overlapping of the strong and weak coupling expansions. We plan to use the density of states to study the 
Fisher's zeros. 
We need numerical confirmation of guesses made for the weak coupling expansion 
for $SU(2)$ where $Im \beta_F/Re \beta_F$ is 5 times larger than for $SU(3)$ .
We need better understanding of  the lattice and the continuum definitions of the gluon condensate.
We need a better understanding of the large order behavior of QCD series in terms of the behavior 
at small complex coupling (a picture analog to metastability  at $\lambda <0$ for the anharmonic oscillator \cite{bender69,Parisi77,brezin77,leguillou90}).

This 
research was supported in part  by the Department of Energy
under Contract No. FG02-91ER40664.


\end{document}